\documentclass[12pt]{article}
\usepackage{epsfig}

\date{}

\newcommand{\be}{\begin{eqnarray}}
\newcommand{\ee}{\end{eqnarray}}
\newcommand{\bpic}{\begin{picture}}
\newcommand{\epic}{\end{picture}}

\begin{document}

\bibliographystyle{unsrt}
\footskip 1.0cm

\thispagestyle{empty}
                           

\vspace{1in}

\begin{center}{\Large \bf {Shadowing of gluons in perturbative 
QCD: ~A comparison of different models}}\\

\vspace{1in}
{\large  Jamal Jalilian-Marian$^1$ and Xin-Nian Wang$^2$}\\

\vspace{.2in}
$^1$ {\it Physics Department, University of Arizona, Tucson, AZ, USA}\\
$^2${\it  Nuclear Science Division, Lawrence Berkeley National Laboratory, 
          Berkeley, CA, USA}\\

\end{center}

\vspace*{25mm}

\begin{abstract}
\baselineskip=18pt
We investigate the different perturbative QCD-based models for nuclear 
shadowing of gluons. We show that in the kinematic region appropriate
to RHIC experiment, all models give similar estimates for the magnitude
of gluon shadowing. At scales relevant to LHC, there is a sizable
difference between predictions of the different models.

\end{abstract}

\vspace*{5mm}

\newpage

\normalsize
\baselineskip=22pt plus 1pt minus 1pt
\parindent=25pt

\section{Introduction}

Understanding the initial stages of a ultra-relativistic heavy 
ion collision is of outmost importance in order to understand the 
outcome of the proposed heavy ion experiments, soon to go online 
at RHIC and later to follow at LHC. Understanding the modifications 
of the parton distributions in nuclei as compared to free nucleons 
(shadowing) will be an important step towards pinning down the 
initial conditions of a heavy ion collision. 

At high energies (small $x$), there are much more gluons than any
other parton species in a hadron/nucleus wavefunction. There are 
a number of processes which are sensitive to gluon shadowing. 
High energy production of mini-jets is one such example.
Mini-jets will be important at RHIC and will dominate at LHC over soft
phenomena. Nuclear shadowing of initial gluon distribution
could significantly reduce the initial mini-jet and total transverse 
energy production. As a result,  the subsequent parton thermalization 
will also be affected due to the reduced initial energy density.
Production of heavy quarks is another example where gluon shadowing
may make a dramatic difference since the probability for making a 
heavy quark pair is proportional to the square of gluon distribution
function and therefore any depletion in number of gluons will make a 
significant difference in the number of heavy quark pairs produced. 

In recent years, there has been considerable progress made towards  
understanding gluon shadowing in perturbative QCD. Shadowing of gluons
defined as
\be
S(x,Q^2,b_t,A)\equiv {xG^A(x,Q^2,b_t) \over AxG^N(x,Q^2,b_t)}
\label{eq:shadow}
\ee
can be understood at high energies as a 
recombination effect due to high gluon number density in the frame 
where the nucleus is fast, the so called Infinite Momentum Frame (IMF) 
or as a multiple scattering effect in the rest frame of the nucleus 
where there is a destructive interference between multiple scattering 
amplitudes. Off course, so long as one calculates the same physical 
quantity, with the same approximations made, one must get the same result.

In this note, we continue our numerical study of the shadowing of 
gluons \cite{jw} using two different QCD based formalisms; one is 
based on an 
all twist, Wilson renormalization group and effective action approach 
to high gluon density region of QCD as developed in \cite{mv,ajmv,rg}. 
This approach takes high gluon densities into effect by 
including and resumming all
$n\rightarrow 1$ ``hard pomeron'' fusion terms in the evolution of both
nuclear and nucleon gluon distribution functions. The other
formalism is based on a generalization of the Mueller-Glauber multiple 
scattering formalism valid in the rest frame of the nucleus  
\cite{mueller,agl,hls,yk} (see also \cite{bal}). The two 
approaches lead to similar but different expressions for the gluon 
distribution function. The difference between the two approaches
is investigated in \cite{kgw} where it is shown that the effective action
and renormalization group approach developed in \cite{mv,ajmv,rg} is
more general than the generalized Mueller-Glauber formalism 
\cite{mueller,agl,hls,yk,bal} and includes effects which are not
present in the latter approach. However, the difference between the
two approaches becomes important only at very high energy (small $x$).
Here we investigate the predictions of the two approaches for gluon 
shadowing numerically and show that the difference between the two is 
negligible 
in the RHIC kinematic region and becomes more appreciable as one goes 
to higher energies. For a review of the experimental status of
nuclear shadowing, we refer the reader to \cite{arn}. For an interesting 
discussion of the role of coherence in nuclear shadowing and its 
manifestation in different frames we refer the reader to \cite{pw}.

This work is organized as follows; in the next section we briefly review
the two formalisms followed by a brief recall of our previous results as 
reported in \cite{jw}. In section $2$, we give the expressions for the
nuclear gluon distribution function from the two different formalisms
and solve the equations numerically and show our results. We finish with 
a discussion.

\section{Shadowing in IMF vs. the rest frame} 

In the infinite momentum frame, shadowing can be understood
as a result of high gluon density at a given impact parameter.
The nucleus is highly Lorentz contracted due to its large speed.
The small $x$ gluons have large wavelengths compared to the
longitudinally Lorentz contracted nucleus size
\be
\lambda \sim {1 \over xp^+} \gg {2R \over \gamma }={2mR \over p^+ }
\nonumber
\ee
Therefore, small $x$ gluons from different nucleons can spatially
overlap and recombine into a higher $x$ gluon. This leads to 
depletion of the nuclear gluon density as compared to naive
expectation that $xG^{A}=AxG^{N}$. In the context of nucleons, this
is referred to as saturation of gluon density and slows down the
unlimited growth of the gluon distribution function which would 
otherwise lead to violation of the unitarity bound on physical
cross sections \cite{fros}.

In \cite{rg}, we derived an evolution equation for the gluon distribution 
function which takes $n \rightarrow 1$ gluon ladder fusion into
account and is the generalization of GLR-MQ model \cite{glrmq}
which includes only $2\rightarrow 1$ ladder recombination.  
The impact parameter dependent evolution equation is 

\be
{\partial^2 \over \partial y \partial \xi}\;xG(x,Q,b_\perp)=
\frac{3}{\pi^3}\ Q^2\bigg[1 - 
{1 \over \kappa} \exp({1\over \kappa}) E_1({1\over \kappa})\bigg]
\label{eq:jklw}
\ee
where $\kappa$ is 
\be
\kappa={N_c \alpha_s \over \pi}{\pi^3 \over 3 Q^2} xG (x,Q,b_\perp)
\ee
with $y=\log 1/x$ and $\xi=\log Q^2$. The exponential integral function
${\rm E_1}(x)$ is \cite{abr}
\be
{\rm E_1}(x)=\int_{0}^\infty\! dt\ {e^{-(1+t)x} \over 1+t},\ \ \ \ x>0.
\label{eq:expint}
\ee 

In \cite{jw}, we numerically solved this equation and calculated
the nucleon and nuclear gluon distribution function at zero
impact parameter. As written, eq. (\ref{eq:jklw}) is a generic
evolution equation for gluons in either nucleons or nuclei. The
distinction between nucleon and nucleus gluon distribution function
is made at the initial point $x_0$ and $Q_0$ after which the nucleon
and nucleus gluon distribution function are determined by the evolution 
equation (see the remarks after eq. \ref{eq:kbar}). We showed in 
\cite{jw} that the non-linearities in the evolution
equation induced by the recombination effects are important. 

In the rest frame of the nucleus, shadowing is manifested through
destructive interference between multiple scattering amplitudes
as described by Glauber-Gribov type models. This approach was used
by Mueller to derive the following expression for the nuclear gluon 
distribution function in perturbative QCD \cite{mueller,agl,hls}
\be
{\partial^2 \over \partial y \partial \xi}\;xG^A(x,Q,b_\perp)=
\frac{2}{\pi^2}\ Q^2\bigg[1 - 
e^{-{1 \over 2} \sigma^{gg}_{N}S(b_t)}\bigg]
\label{eq:mueller}
\ee
where 
\be
\sigma^{gg}_{N} \sim xG^{DGLAP}_N(x,Q^2) 
\label{eq:sigdis}
\ee
is the cross section for scattering of a gluon pair from a nucleon 
inside the nucleus. It is important to realize that relation 
(\ref{eq:sigdis}) holds only at the low gluon density region
and will break down once higher twist effects become important.

This corresponds to the following physical picture in the rest
frame of the nucleus: a highly virtual DIS probe (or a photon) 
fluctuates into a gluon pair (or quark anti-quark pair) well
before it reaches the nucleus. At small $x$, this pair has a long life time
$\tau \sim {1 \over mx}$ and coherently scatters off the nucleons 
as it goes through the nucleus. The destructive interference between
the scattering amplitudes reduces the flux of photons as seen by 
the nucleons sitting inside the nucleus which reduces the nuclear 
cross sections. The Mueller formula (\ref{eq:mueller}) takes 
into account only the interaction of the fastest (or most energetic)
gluon pair with the nucleus. This equation was numerically solved in
\cite{agl,hls} and we refer the reader there for a comparison. 
It was shown in \cite{hls} that Mueller formula
leads to gluon shadowing which is almost independent of the initial 
non-perturbative shadowing input. It is
important to realize that Mueller formula is not a non-linear 
equation for the gluon distribution function the same way GLR-MQ is. For
instance, in Mueller formula, eq. (\ref{eq:sigdis}) reflects
a linear relation between the cross section and the gluon distribution
function. This would not hold were there non-linear effects like higher
twist terms present. It is also known \cite{agl}
that Mueller formula over estimates the amount of shadowing. Ayala 
et al. \cite{agl} proposed to include the effects of scatterings of 
the next-fastest, etc. gluon pairs with the nucleus by iterating the 
Mueller formula. To do this, one replaces the nucleon gluon distribution 
function $xG^N$ in the exponent of eq. (\ref{eq:mueller}) by the nuclear 
gluon distribution function $xG^A$. Furthermore, assuming a Gaussian 
form for the shape function $S(b_t)$, they integrated over the impact 
parameter to get \cite{agl}: 
\be
{\partial^2 \over \partial y \partial \xi}\;xG^A(x,Q)=
\frac{3}{4 \pi^2}\ R^2_A Q^2
\bigg[C + \ln (\kappa_{agl}) + E_1(\kappa_{agl})\bigg] 
\label{eq:agl}
\ee
where $C \sim 0.57$ is the Euler constant and 
\be
\kappa_{agl}=4{N_c \alpha_s \over \pi}{\pi^3 \over 3} 
{1 \over \pi R^2 Q^2} xG^A (x,Q^2)
\label{eq:kagl}
\ee
In \cite{agl}, this replacement is justified by noticing that the new
equation has the correct low density properties such that it reproduces
DLA DGLAP and GLR. It should be emphasized that this modification of
Mueller formula (\ref{eq:mueller}) is just an ansatz which seems to 
produce correct low density limits for the gluon distribution function 
and, unlike the Mueller formula, was not derived from QCD. 
It is shown in \cite{yk} that, with a specific definition of the gluon
distribution function, one can get equation (\ref{eq:agl}) from the $F_2$
structure function calculated from the generalized Mueller-Glauber
formalism. It is interesting that
both approaches predict a slow down in the growth of the gluon
distribution function at very small $x$ such that 
\be
xG(x,Q^2)\sim \pi R^2 Q^2 \ln 1/x.
\label{eq:asymp}
\ee

In the following, in order to compare the two equations, we
will use the same Gaussian ansatz as used by \cite{agl} to integrate
over the impact parameter. We will then solve the two equations
numerically starting with the same exact initial conditions and compare
the results.

\section{Nuclear gluon distribution function}

Here we will use the Gaussian ansatz for the shape function
\be
S(b_t)={1 \over \pi R^2} e^{-{b^2_t \over R^2}}
\label{gauss}
\ee
so that $\int d^2b_t S(b_t)=1$ in order to perform the 
impact parameter integration in eq. (\ref{eq:jklw}). To
do so, we first use the relation (\ref{eq:expint}) to rewrite
the right hand side of eq. (\ref{eq:jklw}) in an integral form.
We then switch the order of integration and perform the impact
parameter integration first and then do the $t$ integration. 
The result is 
\be
{\partial^2 \over \partial y \partial \xi}\;xG(x,Q)=
\frac{3}{\pi^3}\ \pi R^2 Q^2  
\exp({1\over {\bar \kappa}}) E_1({1\over {\bar \kappa}})
\label{eq:intjklw}
\ee
where ${\bar \kappa}$ is 
\be
{\bar \kappa} ={N_c \alpha_s \over \pi}{\pi^3 \over 3}
{1 \over \pi R^2 Q^2} xG (x,Q).
\label{eq:kbar}
\ee 
 
We now solve equations (\ref{eq:intjklw}) and (\ref{eq:agl})
numerically using the methods described in \cite{jw} in full
detail. Here we will just briefly highlight the approximations
made in \cite{jw}. We assumed that at some initial point $x_0$
and $Q_0$, there is no shadowing. This is motivated by experimental 
results \cite{arn,exp} which show that the ratio ${xG^A \over AxG^N}=1$
at $x \sim 0.5-0.6$ with weak $Q^2$ dependence. We then used
the semi-classical approximation to convert our partial differential 
equations into coupled but ordinary differential equations which can
then be solved using Runge-Kutta methods. We refer the reader to 
\cite{jw} for details. The only difference in our choice of parameters in this 
work is the nuclear radius $R$ which was taken to be $5 fm$ for $A=200$
in \cite{jw} while here we use a more realistic value of $7 fm$ 
(more precisely,
$R^A=R_0 A^{1/3}$ with $R_0=1.1 fm$ being the nucleon radius). 
To see the effect of integration over the impact parameter on
shadowing, we first show our result for shadowing at zero impact
parameter in Figure 1.

\begin{figure}[htp]
\centering
\setlength{\epsfxsize=10cm}
\centerline{\epsffile{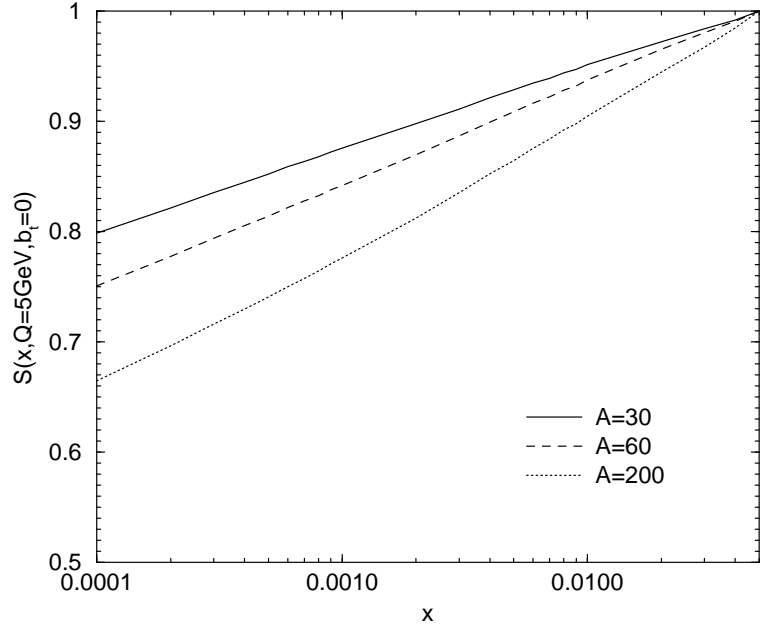}}
\caption{Shadowing from eq. (\ref{eq:jklw}) as a function of $x$ at 
$b_t=0$ and $Q=5 GeV$ for different nuclei.} 
\label{fig:SvsxAbt0}
\end{figure}

The difference between the two equations as a function of $x$
and at fixed $Q=5 GeV$ is shown in Figure 2. Comparing Figures $1$ 
and $2$, it is clear that averaging over the impact parameter
reduces the amount of shadowing as expected. Here, $S^{JKLW}$
refers to the shadowing ratio as defined in (\ref{eq:shadow}) as
calculated from the solution of eq. (\ref{eq:intjklw}) while
$S^{AGL}$ is calculated from the solution to eq. (\ref{eq:agl}).
Also, $S^{GLR}$ is shadowing calculated from the solution of
GLR-MQ eq. \cite{glrmq} which is the second term in the expansion of
eqs. (\ref{eq:intjklw}) and (\ref{eq:agl}). 
The two expressions give very similar results for RHIC ($x\sim 0.01$)
and deviate appreciably only at very small $x$ appropriate to
LHC ($x\sim 0.0001$). For reference, we also show the shadowing ratio 
calculated from GLR-MQ. As expected, GLR-MQ predicts more shadowing 
than the other two expressions.  

\begin{figure}[htp]
\centering
\setlength{\epsfxsize=10cm}
\centerline{\epsffile{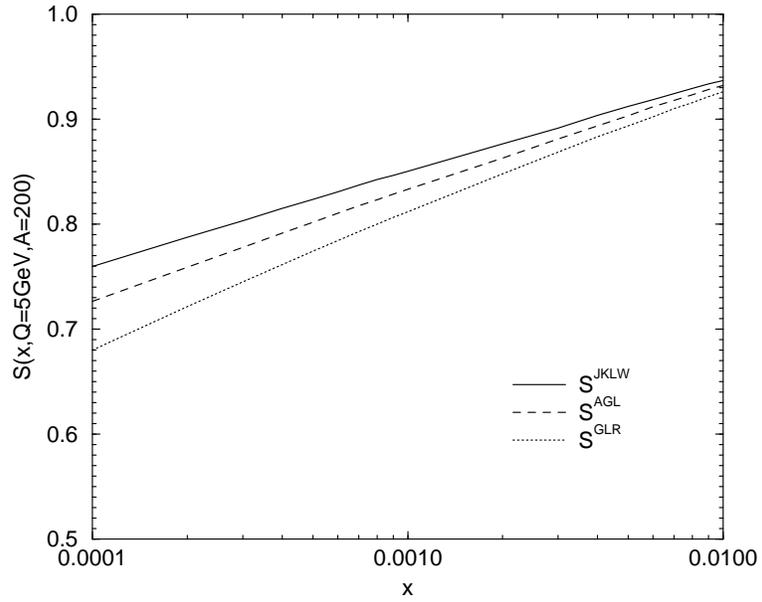}}
\caption{Model dependence of shadowing as a function of $x$ at $Q=5 GeV$.} 
\label{fig:svsxQ5}
\end{figure}

We show the $Q^2$ dependence of shadowing at fixed $x$ 
as predicted by the two equations in Fig. $3$. It is interesting 
to see that both
approaches predict very weak $Q^2$ dependence of gluon shadowing
at RHIC. The difference between the two approaches becomes more 
pronounced at LHC as $x$ gets
smaller. The unphysical trend at low $Q$ is due to both DLA 
and our semi-classical approximation breaking down at small $Q$ 
as discussed in full detail in \cite{jw}.

\begin{figure}[htp]
\centering
\setlength{\epsfxsize=10cm}
\centerline{\epsffile{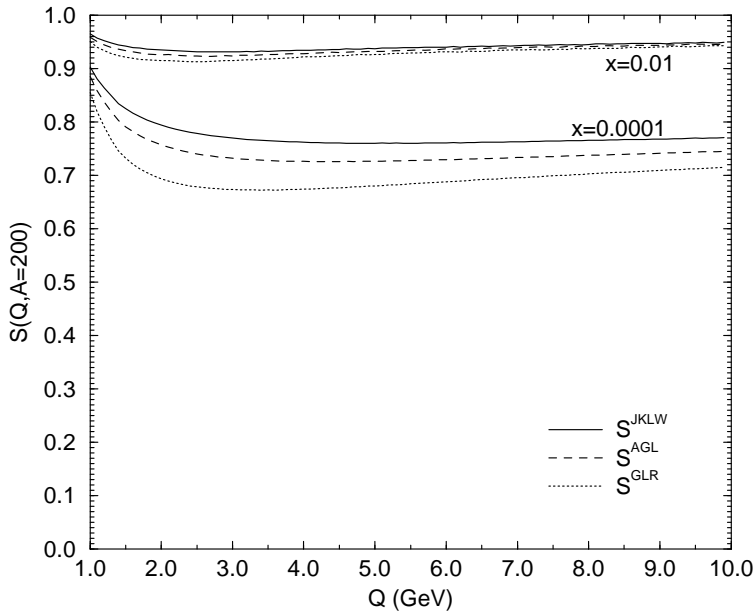}}
\caption{$Q$ dependence of shadowing ratio at RHIC ($x\sim 0.01$) and
LHC ($x\sim 0.0001$).} 
\label{fig:svsQx2x4}
\end{figure}

The dependence of our results on the choice 
of the parameterization chosen for the gluon distribution
function is shown in Figure 4. In \cite{jw}, We used the CTEQ \cite{cteq} 
parameterization to determine the initial gluon distribution 
function at scale $x_0=0.5$ and $Q_0 \sim 1 GeV$. Here we use both
CTEQ and GRV94 \cite{grv94} in order to compare the sensitivity of our
results to the choice of parameterization of gluon distribution function
available. We chose GRV94 since it is the closest in spirit to 
DLA approximation employed in all of the perturbative QCD models. Note 
that GRV94 is known to fail \cite{cald} at small values of $x$ and $Q$ 
but that is not 
relevant here since we use it at a high value of $x$ where it is known 
to work. As seen, the two parameterizations predict similar results for 
RHIC kinematic region while for LHC there is a bigger difference. 

\begin{figure}[htp]
\centering
\setlength{\epsfxsize=10cm}
\centerline{\epsffile{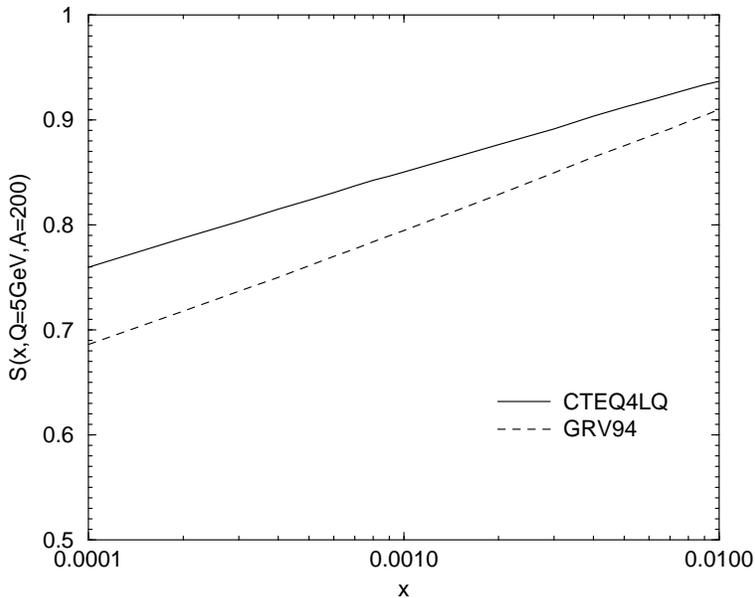}}
\caption{parameterization dependence of shadowing ratio from
eq. (\ref{eq:intjklw}).}
\label{fig:Scteqvsgrv94}
\end{figure}

\section{Discussion}

We discussed the different perturbative QCD-based
models for nuclear gluon distribution function and numerically 
investigated their predictions for $x$ and $Q$ dependence of 
shadowing in the kinematic region appropriate to RHIC and LHC. 
We showed that predictions of different models for shadowing of 
gluons at RHIC are comparable while the difference at LHC can be 
of order $\sim 10\%$ for Gold or Lead. An important point which needs
to be investigated further is inclusion of initial shadowing of gluons
due to non-perturbative effects at the starting point of evolution $x_0$
after which the perturbative evolution takes place. This is currently under
investigation and will be reported on later \cite{jms}.
The more 
experimentally relevant quantity to investigate is the shadowing
of the nuclear structure function $F_2^A$ since shadowing of 
gluons is not directly observable. The all twist $F_2^N$ and
$F_2^A$ as well as the longitudinal structure function $F_L$ were 
computed in \cite{mv} at the classical
level. Including the quantum loop effects due to gluons is straight forward
and is currently under investigation \cite{jmv}. One can then predict
the experimentally measured shadowing ratio $F_2^A/F^2_N$ for different
nuclei at different $x$, $Q^2$ as well as the longitudinal structure
functions.

\leftline{\bf Acknowledgments} 

We would like to thank I. Sarcevic for discussions.
J. J-M. was supported by U.S. Department of Energy under contract No.
DE-FG03-93ER40792. X-N. W. was supported by the Director, Office of 
Energy Research, Office of High Energy and Nuclear Physics Division 
of the Department of Energy, under contract No. DE-AC03-76SF00098 
and DE-FG02-87ER40328.

\leftline{\bf References}

\renewenvironment{thebibliography}[1]
        {\begin{list}{[$\,$\arabic{enumi}$\,$]}  
        {\usecounter{enumi}\setlength{\parsep}{0pt}
         \setlength{\itemsep}{0pt}  \renewcommand{\baselinestretch}{1.2}
         \settowidth
        {\labelwidth}{#1 ~ ~}\sloppy}}{\end{list}}

\end{document}